\documentclass[paper]{ieice}
\usepackage{graphicx,xcolor}
\usepackage[fleqn]{amsmath}
\usepackage{newtxtext}
\usepackage[varg]{newtxmath}
\usepackage{color}

\field{B}
\title{Weyl Spreading Sequence Optimizing CDMA}
\authorlist{%
 \authorentry[tsuda.hirofumi.38u@st.kyoto-u.ac.jp]{Hirofumi Tsuda}{n}{labelA}\MembershipNumber{}
\authorentry[umeno.ken.8z@kyoto-u.ac.jp]{Ken Umeno}{m}{labelA}\MembershipNumber{1101811}
}
\affiliate[labelA]{The author is with the Kyoto University. 
}

\received{2015}{1}{1}
\revised{2015}{1}{1}

\begin{document}
\maketitle
\begin{summary}
This paper shows an optimal spreading sequence in the Weyl sequence class, which is similar to the set of the Oppermann sequences for asynchronous CDMA systems. Sequences in Weyl sequence class have the desired property that the order of cross-correlation is low. Therefore, sequences in the Weyl sequence class are expected to minimize the inter-symbol interference. We evaluate the upper bound of cross-correlation and odd cross-correlation of spreading sequences in the Weyl sequence class and construct the optimization problem: minimize the upper bound of the absolute values of cross-correlation and odd cross-correlation. Since our optimization problem is convex, we can derive the optimal spreading sequences as the global solution of the problem. We show their signal to interference plus noise ratio (SINR) in a special case. From this result, we propose how the initial elements are assigned, that is, how spreading sequences are assigned to each users. In an asynchronous CDMA system, we also numerically compare our spreading sequences with other ones, the Gold codes, the Oppermann sequences, the optimal Chebyshev spreading sequences and the SP sequences in Bit Error Rate. Our spreading sequence, which yields the global solution, has the highest performance among the other spreading sequences tested. 
\end{summary}

\section{Introduction}
%
%
%
%
Signal to Interference plus Noise Ratio (SINR) is an important index for wireless communication systems. In wireless communication systems, it is the most significant parameter in achieving high capacity \cite{shannon}. In general, it is necessary and sufficient for achieving high capacity to increase SINR under the condition that the width of the frequency band is constant when the inter-symbol interference is approximated as Additive White Gaussian Noise (AWGN)\cite{efficiency}. Similarly, the performance of wireless communication is evaluated in Bit Error Rate (BER). However, these two are not independent, and it is known that BER decreases as SINR increases since inter-symbol interference is the most important factor relating to the system performance in CDMA systems.

As a wireless communication system, we focus on a code division multiple access (CDMA) system \cite{dscdma}, in particular, an asynchronous CDMA system. It is one of the multiple access systems with which many people can communicate each other at the same time \cite{multiple}. In CDMA systems, spreading sequences are utilized as codes to multiplex. Each user is assigned a different code and uses it to modulate and demodulate his signal.

In CDMA systems, many methods have been proposed to increase SINR. One such method is based on the blind multi-user detection \cite{mmse}. On the other hand, improving the receiver \cite{ica_cdma} with the application of digital implementation of ICA \cite{ica} and Maximum Likelihood (ML) estimation \cite{ml} is also efficient. However, in particular, ML estimation method needs a large amount of computation.

In general, the value of SINR depends on the spreading sequences. In synchronous CDMA systems, it is known that the Welch bound equality (WBE) sequences realize the maximal capacity \cite{sync}. The Welch bound represents the lower bound of the maximum value of cross-correlation \cite{welch}. When the time delays are given and fixed, the way to find the optimal spreading sequence has been suggested \cite{chipsync}. However, in general asynchronous CDMA systems, the optimal spreading sequences have not been found, and asynchronous CDMA systems have been investigated \cite{fundamental}.

In uplink of W-CDMA systems, the current spreading sequence is the Gold code \cite{gold}. It is known \cite{prop} that the Gold code is optimal in all the binary spreading sequences as well as the Kasami sequence \cite{kasami} in a sense of the maximum value among all periodic autocorrelation and periodic cross-correlation. To explore a better sequence for asynchronous CDMA systems, in \cite{chaos_cdma} and \cite{chaos_mod}, the use of chaotic spreading sequences has been proposed. These chaotic spreading sequences are multivalued sequences, not binary ones, and are obtained from chaotic maps. Examples of such spreading sequences have been given in \cite{logistic}-\cite{mazzini}. In \cite{limit}, the approach to obtain the capacity of spreading sequences has been proposed.

In \cite{sarwate}, Sarwate has shown two kinds of characterized sequences on his limitation. One kind is a set of sequences whose periodic cross-correlation is always zero. We call them Sarwate sequences. The other kind is a set of sequences whose periodic autocorrelation is always zero except for only one point, that is, Frank-Zadoff-Chu (FZC) sequences \cite{zadoff} \cite{chu}. In \cite{FZC_fam}, the extended set of the FZC sequences, the Oppermann sequences are proposed. They have three parameters and their SINR, autocorrelation and cross-correlation have been investigated.

In this paper, we define the Weyl sequence class, which is a set of sequences generated by the Weyl transformation \cite{weyl}. This class is similar to the set of Oppermann sequences and includes the Sarwate sequences. Sequence in the Weyl sequence class have a desired property that the order of cross-correlation is low. We evaluate the upper bound of cross-correlation and construct the optimization problem: minimize the upper bound of cross-correlation. From the problem, we derive optimal spreading sequences in the Weyl sequence class. We show their SINR in a special case and compare them with other sequences in a sense of BER.

\section{Weyl Sequence Class}
In this section, we define the Weyl sequence class and show their properties. Let $N$ be the length of spreading sequences. We define the Weyl sequence $(x_n)$ as the following formula \cite{weyl}
\begin{equation}
x_n = (n \rho + \Delta) \bmod 1 \hspace{4mm}(n=1,2, \ldots, N),
\end{equation}
where $\rho$ and $\Delta$ are real parameters. From the above definition, we can assume that the parameters $\rho$ and $\Delta$ satisfy $0 \leq \Delta < 1$ and $0 \leq \rho < 1$. The sequences whose $\rho$ is an irrational number are used in a Quasi-Monte Carlo method \cite{dp}. We apply this sequence to a spreading sequence. Then, the Weyl spreading sequence $(w_{k,n})$ is defined as \cite{almost}
\begin{equation}
\begin{split}
x_{k,n} &= (n\rho_k + \Delta_k) \bmod 1,\\
w_{k,n} &= \exp(2 \pi j x_{k,n}) \hspace{4mm}(n=1,2, \ldots, N),
\end{split}
\end{equation}
where $k$ is the number of the user, $j$ is the unit imaginary number, and $\rho_k$ is a real-valued initial point assigned to the user $k$. In CDMA systems, the value of $\Delta_k$ has no effects to Signal to Interference plus Noise Ratio (SINR) since $\exp(2 \pi j \Delta_k)$ is united to the phase term of the signal. Thus, we set $\Delta_k = 0$. We call the class which consists of Weyl spreading sequences the Weyl sequence class. Note that this class is similar to the set of Oppermann sequences \cite{FZC_fam}. The $n$-th element of the Oppermann sequences is defined as
\mathindent=0mm
\begin{equation}
u_{k,n} = (-1)^{nM_k}\exp\left(\frac{j\pi (M_k^pn^q+n^r)}{N}\right)\hspace{2mm}(n=1,2, \ldots, N),
\label{eq:opper}
\end{equation}
\mathindent=7mm
where $M_k$ is an integer that is relatively prime to $N$ such that $1 \leq M_k < N$ and $p, q$ and $r$ are any real numbers. The triple $\{p,q,r\}$ specifies the set of sequences. When the triple $\{p,q,r\}$ is $\{2,1,-\infty\}$, we obtain the element of the FZC sequence \cite{zadoff}\cite{chu}
\begin{equation}
u_{k,n} = (-1)^{nM_k}\exp\left(\frac{j\pi (M_k^2n)}{N}\right).
\end{equation}
The Weyl sequence class is obtained when the triple is $\{1,1,-\infty\}$ and $M_k = \rho_k \cdot 2N/(N+1)$. The number $-1$ is written as $\exp(j\pi)$. Substituting $\{p,q,r\}=\{1,1,-\infty\}$ and $M_k = \rho_k \cdot 2N/(N+1)$ into Eq. (\ref{eq:opper}), we have
\begin{equation}
\begin{split}
u_{k,n} &= \exp(j\pi)^{nM_k}\exp\left(\frac{j\pi (M_k^pn^q+n^r)}{N}\right)\\
&= \exp\left(2\pi j n\frac{N}{N+1}\rho_k\right)\exp\left(2\pi j n\frac{1}{N+1}\rho_k\right)\\
&= \exp\left(2\pi j n\rho_k\right).
\end{split}
\label{eq:weyl_triple}
\end{equation}
Note that $M_k$ is not always an integer. Thus, the Weyl sequence class is similar to the set of Oppermann sequences.

The element of the Weyl sequence class, $(w_{k,n})$ has a desired property that cross-correlation is low. We define the periodic correlation function $\theta_{i,k}(l)$ and odd periodic correlation function $\hat{\theta}_{i,k}(l)$ as
\begin{eqnarray}
\theta_{i,k}(l) &=& C_{i,k}(l) + C_{i,k}(l-N), \\
\hat{\theta}_{i,k}(l)&=& C_{i,k}(l) - C_{i,k}(l-N),
\end{eqnarray}
where
\begin{equation}
C_{i,k}(l) = \left\{ 
\begin{array}{c c}
\displaystyle \sum_{n=1}^{N-l}\overline{w_{i,n+l}} w_{k,n}  & 0 \leq l \leq N-1,\\
\displaystyle \sum_{n=1}^{N+l} \overline{w_{i,n}}  w_{k,n-l} & 1-N \leq l < 0,\\
0 & | l | \geq N\\
\end{array} \right. 
\end{equation}
and $\overline{z}$ is the conjugate of $z$.
The correlation functions $\theta_{i,k}(l)$ and $\hat{\theta}_{i,k}(l)$ have been studied in \cite{welch} \cite{sarwate} \cite{cross}. When $i \neq k$, $\theta_{i,k}(l)$ and $\hat{\theta}_{i,k}(l)$ are the periodic function and the odd periodic cross-correlation function, respectively. It is necessary for achieving high SINR to keep the value of the cross-correlation functions, $|\theta_{i,k}(l)|$ and $|\hat{\theta}_{i,k}(l)|$ low for all $0 \leq l < N $. The absolute values of cross-correlation functions $|\theta_{i,k}(l)|$ and $|\hat{\theta}_{i,k}(l)|$ have the common upper bound that
\begin{eqnarray}
|\theta_{i,k}(l)| &\leq& |C_{i,k}(l)| + |C_{i,k}(l-N)|, \\
|\hat{\theta}_{i,k}(l)|&\leq& |C_{i,k}(l)| + |C_{i,k}(l-N)|.
\end{eqnarray}
With the sequences in the Weyl sequence class, we evaluate the absolute value of $C_{i,k}(l)$ as
\begin{equation}
\begin{split}
|C_{i,k}(l)| &= \left|\frac{1 - \exp(2 \pi  j (N-l)(\rho_k - \rho_i))}{1 - \exp(2 \pi j (\rho_k - \rho_i))}\right|\\
&=\sqrt{\frac{1 - \cos(2  \pi (N-l)(\rho_k - \rho_i))}{1 - \cos(2 \pi (\rho_k - \rho_i))}} \\
& =  \left|\frac{\sin( \pi (N-l) (\rho_k - \rho_i))}{\sin(\pi (\rho_k - \rho_i))}\right| \\
& \leq  \frac{1}{|\sin(\pi (\rho_k - \rho_i))|} \\
& = \frac{1}{|\sin(\pi (\rho_i - \rho_k))|}.
\end{split}
\label{eq:bound1}
\end{equation}
The equality is attained if and only if 
\begin{equation}
(N-l)(\rho_k - \rho_i) = \frac{1}{2} + m,
\end{equation}
where $m \in \mathbb{Z}$. 
From the above result, $|C_{i,k}(l)|$ obeys
\begin{equation}
|C_{i,k}(l)| = O\left(1\right),
\label{eq:prop}
\end{equation}
with $O$ being an order function.
Similarly, $|C_{i,k}(l-N)|$ obeys $O\left(1\right)$.
Thus, the common upper bound of $|\theta_{i,k}(l)|$ and $|\hat{\theta}_{i,k}(l)|$ is independent of $N$. For general spreading sequences, due to the central limit theorem (CLT), the cross-correlations $|\theta_{i,k}(l)|$ and $|\hat{\theta}_{i,k}(l)|$ become large as $N$ becomes large. For this reason, compared to general spreading sequences, the Weyl spreading sequence is expected to have low cross-correlation. 

\section{Optimal Spreading Sequence in Weyl Sequence Class}
In this section, we consider an asynchronous binary phase shift keying (BPSK) CDMA system. Our goal is to derive the spreading sequences whose inter-symbol interference is the smallest in the Weyl sequence class. Let $K$, $T$ and $T_c$ be the number of users, the durations of the symbol and each chip, respectively. In this situation, the user $i$ despreads the spreading sequences $(w_{k,n})$ with the spreading sequence of the user $i$, $(w_{i,n})$. The symbols $b_{k,-1}, b_{k,0} \in \{-1,1\}$ denote bits which the user $k$ send. The transmitted signal of the user $k$ has time delay $\tau_k$. From \cite{pursley}, we assume that time delay $\tau_k$ is distributed in $[0,T)$ and satisfies $l_kT_c \leq \tau_k \leq (l_k + 1)T_c$, where $l_k \in \{0,1,\ldots,N-1\}$ is an integer. Then, the inter-symbol interference between the user $i$ and the user $k$, $I_{i,k}(\tau_k)$ is obtained as
\mathindent=0mm
\begin{equation}
\begin{split}
I_{i,k}(\tau_k) &= \exp(j\phi_k)\left[(\tau_k - l_kT_c)\left(b_{k,-1}C_{i,k}(l_k) \right.\right.\\
 &\left. \left. + b_{k,0}C_{i,k}(l_k-N) \right) \right. + ((l_k+1)T_c - \tau_k)\\
&\left. \cdot \left\{b_{k,-1}C_{i,k}(l_k+1) + b_{k,0}C_{i,k}(l_k+1-N) \right\} \right], 
\label{eq:correlation}
\end{split}
\end{equation}
\mathindent=7mm
where $\phi_k \in [0,2\pi)$ is the phase of user $k$'s carrier. With Eq. (\ref{eq:bound1}), the absolute value of the inter-symbol interference $I_{i,k}(\tau_k)$ is evaluated as
\begin{equation}
\begin{split}
|I_{i,k}(\tau_k)| \leq& (\tau_k - l_kT_c)\{|C_{i,k}(l_k)| + |C_{i,k}(l_k-N)| \}\\
+& ((l_k+1)T_c - \tau_k)\\
\cdot&\{|C_{i,k}(l_k+1)| + |C_{i,k}(l_k+1-N)| \}\\
\leq &\frac{2T_c}{|\sin(\pi (\rho_i - \rho_k))|}.
\end{split}
\label{eq:bound2}
\end{equation}
Thus, we have shown that the upper bound of inter-symbol interference between two sequences is inversely proportional to $|\sin(\pi d(\rho_i - \rho_k))|$.
To reduce the inter-symbol interference $I_{i,k}(\tau_k)$, it is necessary to reduce $2T_c/|\sin(\pi (\rho_i - \rho_k))|$.
To eliminate the absolute value function, we introduce the distance between the phases $\rho_i$ and $\rho_k$. The distance $d(\rho_i,\rho_k)$ we propose here is given by
\begin{equation}
d(\rho_i,\rho_k) = \min\{|\rho_i - \rho_k|,1 - |\rho_i - \rho_k|\}.
\label{eq:d}
\end{equation}
Note that this $d$ satisfies the axiom of distance, and
\begin{equation}
|\sin(\pi (\rho_i - \rho_k))| = \sin(\pi d(\rho_i,\rho_k)),
\label{eq:d_prop1}
\end{equation}
\begin{equation}
0 \leq d(\rho_i,\rho_k) \leq \frac{1}{2}
\label{eq:d_prop2}
\end{equation}
if we regard $\rho=1$ in the same light as $\rho=0$.
From Eq. (\ref{eq:d}), we rewrite Eq. (\ref{eq:bound2}) without any absolute value as
\begin{equation}
|I_{i,k}(\tau_k)| \leq \frac{2T_c}{\sin(\pi d(\rho_i, \rho_k))}.
\label{eq:bound3}
\end{equation}
We should take into account the whole inter-symbol interference in the users. The whole inter-symbol interference $I$ is written as
\begin{equation}
I = \sum_{i=1}^K \sum_{\substack{k=1 \\ k \neq i}}^K I_{i,k}.
\end{equation}
With Eq. (\ref{eq:bound3}), it is clear that $|I|$ has the upper bound:
\begin{equation}
|I| \leq \sum_{i=1}^K \sum_{\substack{k=1 \\ k \neq i}}^K  \frac{2T_c}{\sin(\pi d(\rho_i, \rho_k))}.
\label{eq:whole}
\end{equation}
Thus, we minimize Eq. (\ref{eq:whole}) and obtain the problem $(\tilde{P})$
\begin{equation}
\begin{split}
(\tilde{P})&\hspace{3mm} \min \hspace{3mm} \sum_{i=1} \sum_{\substack{k=1 \\ k \neq i}} \frac{1}{\sin(\pi d(\rho_i , \rho_k))}\\
\mbox{subject to} & \hspace{3mm} \rho_k \in [0,1) \hspace{2mm}(1 \leq k \leq K).
\end{split}
\end{equation}
This problem is equivalent to that we minimize the sum of the upper bound of $C_{i,k}(l)$. Thus, the cross-correlation among all the users is expected to be always low when we solve this problem. From Eq. (\ref{eq:d_prop1}), it is clear that $d(\rho_i,\rho_k) = d(\rho_k,\rho_i)$. Then, in the problem $(\tilde{P})$, we count two times the same distance. Thus, we obtain the equivalent problem $(\tilde{P'})$
\begin{equation}
\begin{split}
(\tilde{P'})&\hspace{3mm} \min \hspace{3mm} \sum_{i<k} \frac{1}{\sin(\pi d(\rho_i , \rho_k))}\\
\mbox{subject to} & \hspace{3mm} \rho_k \in [0,1) \hspace{2mm}(1 \leq k \leq K).
\end{split}
\end{equation}
It is not clear whether the objective function of the problem $(\tilde{P'})$ is convex or not since the form of function $d$ is complicated. To eliminate the function $d$, we introduce slack variables $t_{i,k}$ for $(P)$. Then, the problem $(\tilde{P'})$ is rewritten as
\begin{equation}
\begin{split}
(\tilde{P''}) &\hspace{3mm} \min \hspace{3mm}\sum_{i<k} \frac{1}{\sin(\pi t_{i,k})},\\
\mbox{subject to}  & \hspace{3mm} \rho_k \in [0,1) \hspace{2mm} (1 \leq k \leq K),\\
& \hspace{3mm} |\rho_i - \rho_k | \geq t_{i,k} \hspace{2mm} (i<k), \\
& \hspace{3mm} 1 - |\rho_i - \rho_k | \geq t_{i,k}\hspace{2mm} (i<k), \\
& \hspace{3mm} t_{i,k} \geq 0\hspace{2mm} (i<k).
\end{split}
\end{equation}
Without loss of generality, we assume $\rho_k \leq \rho_{k+1}$. Then, the problem $(\tilde{P''})$ can be rewritten as
\begin{equation}
\begin{split}
(P)& \hspace{3mm}\min \hspace{3mm} \sum_{i<k} \frac{1}{\sin(\pi t_{i,k})}, \\
\mbox{subject to} &\hspace{3mm}   \rho_k - \rho_i \geq t_{i,k} \hspace{2mm} (i<k), \\
&\hspace{3mm}  1 - \rho_k + \rho_i\geq t_{i,k}\hspace{2mm} (i<k), \\
&\hspace{3mm}  \rho_{i+1} \geq \rho_i \hspace{2mm} (1\leq i \leq K-1), \\
&\hspace{3mm}  \rho_1 \geq 0 , \rho_K \leq 1, \\
&\hspace{3mm} t_{i,k} \geq 0\hspace{2mm} (i<k).
\end{split}
\end{equation}
Notice that the objective function and the inequality constraints of the problem $(P)$ are convex. It has been known that convex programming can be solved with the KKT conditions \cite{KKT}. To write such conditions, we define the variable vector $\mathbf{x}$ as
\begin{equation}
\begin{split}
\boldsymbol{\rho} =& (\rho_1,\rho_2,\ldots,\rho_K)^\mathrm{T}, \\
\mathbf{t} =& (t_{1,2},t_{1,3},\ldots,t_{1,K},t_{2,3},\ldots,t_{K-1,K})^\mathrm{T}, \\
\mathbf{x}=& \left( \begin{array}{c}
\boldsymbol{\rho} \\
\mathbf{t}\\
\end{array} \right), 
\end{split}
\end{equation}
where $\boldsymbol{\rho} \in \mathbb{R}^K$, $\mathbf{t} \in \mathbb{R}^{K(K-1)/2}$, $\mathbf{x} \in \mathbb{R}^{K(K+1)/2}$ and $\mathbf{z}^\mathrm{T}$ is the transpose of $\mathbf{z}$. From the KKT conditions, the solution $\mathbf{x}^*$ is a global solution of $(P)$ if $\mathbf{x}^*$ satisfies the following equation:
\mathindent=0mm
 \begin{equation}
\begin{split}
&\nabla f(\mathbf{x}^*) + \sum_{i<k}\lambda_{i,k}\nabla c_{i,k}(\mathbf{x}^*) + \sum_{i<k}\mu_{i,k}\nabla d_{i,k}(\mathbf{x}^*)\\
 +& \sum_{i=1}^{K-1}\nu_i \nabla e_i(\mathbf{x}^*) +\xi_1\nabla g_1(\mathbf{x}^*)\\
 +& \xi_K\nabla g_K(\mathbf{x}^*) + \sum_{i<k}o_{i,k}\nabla h_{i,k}(\mathbf{x}^*) =  \mathbf{0},
\end{split}
\label{eq:KKT1_1}
 \end{equation}
 \mathindent=7mm
where
\begin{equation}
\begin{split}
f(\mathbf{x}) &= \sum_{i<k} \frac{1}{\sin(\pi t_{i,k})}, \\
c_{i,k}(\mathbf{x}) &= t_{i,k} + \rho_i - \rho_k,\\
d_{i,k}(\mathbf{x}) &= t_{i,k} -1 - \rho_i + \rho_k,\\
e_i(\mathbf{x}) &= \rho_i - \rho_{i+1},\\
g_1(\mathbf{x}) &= -x_1,\\
g_K(\mathbf{x}) &= x_K -1, \\
h_{i,k}(\mathbf{x}) &= -t_{i,k},
\end{split}
\label{eq:KKT1_2}
\end{equation}
and the Lagrange multipliers $\lambda_{i,k}, \mu_{i,k}, \nu_{i}, \xi_{1}, \xi_{K}$ and $o_{i,k}$ are non-negative real numbers. They have to satisfy the following conditions:
\begin{equation}
\begin{split}
c_{i,k}(\mathbf{x}) < 0 &\Rightarrow \lambda_{i,k} = 0,\\
d_{i,k}(\mathbf{x}) < 0 &\Rightarrow \mu_{i,k} = 0,\\
e_{i}(\mathbf{x}) < 0 &\Rightarrow \nu_{i} = 0,\\
g_{1}(\mathbf{x}) < 0 &\Rightarrow \xi_{1} = 0,\\
g_{K}(\mathbf{x}) < 0 &\Rightarrow \xi_{K} = 0,\\
h_{i,k}(\mathbf{x}) < 0 &\Rightarrow o_{i,k} = 0.
\end{split}
\label{eq:KKT1_3}
\end{equation}
In the appendix A, we prove that the global optimal solutions $\rho^*_i$ and $t^*_{i,k}$ are given by
\begin{equation}
\begin{split}
\rho^*_i &= \gamma + \frac{i-1}{K} \hspace{2mm}(i=1,2,\ldots,K),\\
t^*_{i,k} &= \min\left\{\frac{|k - i|}{K},1-\frac{|k - i|}{K}\right\},
\end{split}
\label{eq:global_solution}
\end{equation}
where $\gamma$ is a real number. From the above result, the optimal spreading sequence of the user $k$, $(\tilde{w}_{k,n})$ is given by
\begin{equation}
\tilde{w}_{k,n} = \exp\left(2 \pi j n\left(\gamma + \frac{k-1}{K}\right)\right)
\end{equation}
for a real number $\gamma$. This is equivalent to the following spreading sequences:
\begin{equation}
\tilde{w}_{k,n} = \exp\left(2 \pi j n\left(\gamma + \frac{\sigma_k}{K}\right)\right),
\label{eq:optimal_seq}
\end{equation}
where $\sigma_k \in \{0,1,2,\ldots,K-1\}$ which satisfies $\sigma_k \neq \sigma_i$ when $k \neq i$. This sequence belongs to the Weyl sequence class. Therefore, similar to Eq. (\ref{eq:weyl_triple}), this sequence is obtained from the triple $\{1,1,-\infty\}$ and $M_k=2N(\gamma + \sigma_k/K)/(N+1)$ in the Oppermann sequences. If $N=2(L+1)$, $K=L+1$ and $\gamma = \frac{1}{2}$, where $L$ is an even number, then our sequences are equivalent to the Song-Park (SP) sequences ($\sigma_k = L+1$ is not used) \cite{song_park}. If $K=N$  and $\gamma=0$,  our sequences are equivalent to the Sarwate sequences \cite{sarwate}. 

\section{SINR with the Optimal Spreading Sequence}
In the previous section, we fix the number of users $K$ and derive optimal spreading sequences in the Weyl spreading sequence class. This spreading sequence is expected to be useful when the number of the users $K$ is fixed. However, the number of users in a channel changes as time passes. Thus, in this section, we fix the maximum number of users in a channel, $K_{\max}$ and assign the $\sigma_k \in \{0,1,2,\ldots,K_{\max}-1\}$ to $K$ users.

The spreading sequences assigned to the user $k$ is expressed as
\mathindent=0mm
\begin{equation}
\tilde{w}_{k,n}(K_{\max},\gamma) = \exp\left(2 \pi j n\left(\gamma + \frac{\sigma_k}{K_{\max}}\right)\right)\hspace{2mm}(n=1,2,\ldots,N).
\label{eq:suggest}
\end{equation}
\mathindent=7mm
Note that the optimal spreading sequence of the problem ($P$) is expressed as
\begin{equation}
\tilde{w}_{k,n}(K,\gamma) = \exp\left(2 \pi j n\left(\gamma + \frac{\sigma_k}{K}\right)\right)\hspace{2mm}(n=1,2,\ldots,N).
\end{equation}
In particular, when $K_{\max} = N$ and $\gamma = 0$, this spreading sequence is equivalent to the Sarwate's sequence \cite{sarwate}. The Sarwate's sequence has the feature that the periodic cross-correlation function is always 0, that is,
\begin{equation}
\theta_{i,k}(l) = C_{i,k}(l) + C_{i,k}(l-N) = 0
\label{eq:cor_sarwate}
\end{equation}
for all $l$ and $k \neq i$.

When we set $K_{\max} = N$, $\theta_{i,k}(0)=0$ for all the $\gamma$ and $k \neq i$. Thus, the sequence $\tilde{w}_{k,n}(N,\gamma)$ is the WBE sequence since the orthogonal condition is satisfied. From the above reason, we define $K_{\max}$ as $N$, that is, we consider the following sequences
\mathindent=0mm
\begin{equation}
\tilde{w}_{k,n}(N,\gamma) = \exp\left(2 \pi j n\left(\gamma + \frac{\sigma_k}{N}\right)\right)\hspace{2mm}(n=1,2,\ldots,N).
\label{eq:sarwate_seq}
\end{equation}
\mathindent=7mm
In this section, we assume that $\sigma_k$ is a random variable and is uniformly distributed in $\{0,1,2, \ldots, N-1\}$. In the next section, we consider how to assign $\sigma_k$ in a systematic approach.

The expression of SINR of the user $i$ is obtained in \cite{pursley} \cite{mazzini} as
\begin{equation}
\mbox{SINR}_i = \left\{(6N^3)^{-1} \sum_{\substack{ k=1\\ k \neq i}}^{K}r_{i,k} + \frac{N_0}{2E} \right\}^{-1/2},
\end{equation}
where
\mathindent=0mm
\begin{equation}
\begin{split}
r_{i,k} =& \sum_{l=0}^{N-1} \{ |C_{i,k}(l-N)|^2 + \operatorname{Re}[C_{i,k}(l-N) \overline{C_{i,k}(l-N-1)}] \\
 &+ |C_{i,k}(l-N+1)|^2 + |C_{i,k}(l)|^2 \\
& + \operatorname{Re}[C_{i,k}(l) \overline{C_{i,k}(l+1)}] +  |C_{i,k}(l+1)|^2 \},
\end{split}
\label{eq:SNR}
\end{equation}
\mathindent=7mm
$E$ is the energy per data bit and $N_0$ is the power of Gaussian noise. SINR is the ratio between the variance of a desired signal and the one of a noise signal. 
In the appendix B, with the spreading sequence $\{\tilde{w}_{k,n}(N,\gamma)\}$, we prove that SINR of the user $i$ is given by
\begin{equation}
\mbox{SINR}_i=\left\{ R_i +  \frac{N_0}{2E} \right\}^{-1/2},
\label{eq:upper}
\end{equation}
where
\mathindent=0mm
\begin{equation}
R_i = \frac{(K-1)}{18N^2}\left\{2(N+1) +\left(N-2\right)\cos\left(2 \pi \left(\gamma + \frac{\sigma_i}{N}\right)\right)\right\}.
\label{eq:upper_sub}
\end{equation}
\mathindent=7mm
Equation (\ref{eq:upper}) is obtained when the ratio $K/N$ is close to $1$, that is, the number of users $K$ is sufficiently large. From Eqs. (\ref{eq:upper}) and (\ref{eq:upper_sub}), the spreading sequence $\{\tilde{w}_{i,n}(N,\gamma)\}$ has different SINR in $\sigma_i$. Thus, some users have high SINR and other users have low SINR. The lower bound of $\mbox{SINR}_i$ is
\begin{equation}
\underline{\mbox{SINR}}_i = \left\{\frac{K-1}{6N} + \frac{N_0}{2E} \right\}^{-\frac{1}{2}}\label{eq:snr_upper}.
\end{equation}

\section{How to Assign $\sigma_k$}
In this section, we consider how to assign $\sigma_k$ to the each users. Let us consider the spreading sequences
\[\tilde{w}_{k,n}(N,\gamma) = \exp\left(2\pi j n\left(\gamma + \frac{\sigma_k}{N}\right)\right),\]
where $\sigma_k \in \{0,1,\ldots,N-1\}$. From Eq. (\ref{eq:global_solution}), it is demanded that we assign $\sigma_k$ at regular interval. However, these sequences cannot be used if the number of users changes. Thus, we have to make the rule to assign $\sigma_k$ when the number of users changes.

We give the rule to assign $\sigma_k$ in the situation that the number of users monotonic increases. From the demand of the problem $(P)$, it is desirable that we assign $\sigma_k$ to each users at regular interval. Thus, it is appropriate to assign them at nearly regular interval in every number of users. We apply the Van der Corput sequence \cite{vandercorput} to the method to assign $\sigma_k$ since the sequence is a regular interval sequence in some situations. For example, the Van der Corput sequence $(v_n)$ is obtained as
\begin{equation}
(v_n)=\left\{0,\frac{1}{2},\frac{1}{4},\frac{3}{4},\frac{1}{8},\frac{5}{8},\frac{3}{8}, \frac{7}{8}, \frac{1}{16},\ldots\right\}.
\end{equation}
In particular, when we take the first eight elements out from $(v_n)$ and sort them, we obtain the sequence
\begin{equation}
\left\{0,\frac{1}{8},\frac{2}{8},\frac{3}{8},\frac{4}{8},\frac{5}{8},\frac{6}{8}, \frac{7}{8}\right\}.
\end{equation}
This sequence is a regular interval sequence. We can consider $(v_n)$ as a nearly regular interval sequence.

When the length of spreading sequences $N$ equals $2^m$, where $m > 1$ is an integer, $(v_n)$ is rewritten in terms of $1/N$. For example, when $N=16$, the sequence $(v_n)$ is obtained as
\begin{equation}
(v_n)=\left\{0,\frac{8}{16},\frac{4}{16},\frac{12}{16},\frac{2}{16},\frac{10}{16},\frac{6}{16}, \frac{14}{16}, \frac{1}{16},\ldots\right\}.
\end{equation}
Thus, we propose that we use the $k$-th element of $(v_n)$ as $\sigma_k/N$, that is, the spreading sequences are expressed as
\begin{equation}
\tilde{w}_{k,n}(N) = \exp\left(2 \pi j n \left(\gamma + v_k \right) \right),
\label{eq:vander_seq}
\end{equation}
where $v_k$ is the $k$-th element of $(v_n)$.

\section{Simulation Results}
In this section, we simulate an asynchronous CDMA system and discuss the performance of the spreading sequences obtained by Eqs. (\ref{eq:optimal_seq}) (\ref{eq:sarwate_seq}) (\ref{eq:vander_seq}). We use two parameters $\gamma = 1/(2N)$ and $\gamma = 1/(2K)$. Choosing the parameter $\gamma=0$, we have a spreading sequence whose elements are $1$. This sequence is trivial. Therefore, to avoid such a sequence, we use the two cases for $\gamma$, $\gamma = \frac{1}{2N}$ and $\gamma = \frac{1}{2K}$. We focus on BPSK and QPSK systems with AWGN channel and no fading signals. In this simulation, we make the following assumptions about the receivers and the channel
\begin{enumerate}
\item the receiver has the perfect synchronization with the desired signal and no knowledge about the time delay of the other signals.
\item there are no fading effects.
\item the time delay $\tau_k$,  the symbols $b_{k,-1}$ and $b_{k,0}$, and the phase $\theta_k$ are normally distributed in $[0,T)$, $\{-1,1\}$, and $[0,2\pi)$, where $T$ is the duration of each symbol.
\item the spreading sequences are uniformly and randomly chosen. With the Weyl spreading sequences, the parameter $\sigma_k$ is uniformly and randomly chosen.
\item the matched filter is used in the correlation receiver.
\end{enumerate}

The detail of transmitters and receivers are described in \cite{ergotic} \cite{mazzini} \cite{pursley}.

We measure the averaged BER
\begin{equation}
\mbox{BER} = \frac{1}{K}\frac{1}{U}\sum_{k=1}^K \sum_{u=1}^U \mbox{BER}_{k,u},
\end{equation}
where $U$ is the trial numbers, $u$ is the $u$-th trial number and $\mbox{BER}_{k,u}$ is the BER of the user $k$ at the $u$-th trial. This section consists of three subsections. In the first subsection, we compare the spreading sequences obtained by Eq. (\ref{eq:optimal_seq}) and (\ref{eq:sarwate_seq}) with other sequences, the Gold codes \cite{gold}, the optimal Chebyshev spreading sequences \cite{ergotic} and the Oppermann sequences \cite{FZC_fam}. In particular, we choose the triple $\{p,q,r\} = \{1.0,1.0,1.275\}$ as the parameters of the Oppermann sequences. This triple is shown in \cite{FZC_fam} as the optimal parameters when $N=31$ with $N$ being the length. In the second subsection, we compare the spreading sequences obtained by Eq. (\ref{eq:optimal_seq}) and (\ref{eq:sarwate_seq}) with one obtained by Eq. (\ref{eq:vander_seq}). We compare the random assigning approach with the systematic approach. In the final subsection, we compare the spreading sequences obtained by Eq. (\ref{eq:optimal_seq}) and (\ref{eq:sarwate_seq}) with the SP sequences \cite{song_park}. 

\subsection{Comparison with Other Sequences}
We consider the following spreading sequences:
\mathindent=0mm
\begin{equation*}
\begin{split}
\tilde{w}_{k,n}(N,\gamma) &= \exp\left(2 \pi j n\left(\gamma + \frac{\sigma_k}{N}\right)\right)\hspace{4mm}(n=1,2,\ldots,N),\\
\tilde{w}_{k,n}(K,\gamma) &= \exp\left(2 \pi j n\left(\gamma + \frac{\sigma_k}{K}\right)\right)\hspace{4mm}(n=1,2,\ldots,N).
\end{split}
\end{equation*}
\mathindent=7mm
The former sequence is obtained from Eq. (\ref{eq:sarwate_seq}) and the latter sequence is obtained from Eq. (\ref{eq:optimal_seq}). In this section, the ``Weyl'' spreading sequence is $\left\{\tilde{w}_{k,n}(N,\gamma)\right\}$ and the ``Optimal'' one is $\left\{\tilde{w}_{k,n}(K,\gamma)\right\}$. Note that the optimal spreading sequences are different in the number of users $K$. The ``Upper Bound'' is obtained from Eq. (\ref{eq:snr_upper}).

\begin{figure*}[htb] 
\begin{minipage}{0.50\hsize}
   \centering
   \includegraphics[width=3in]{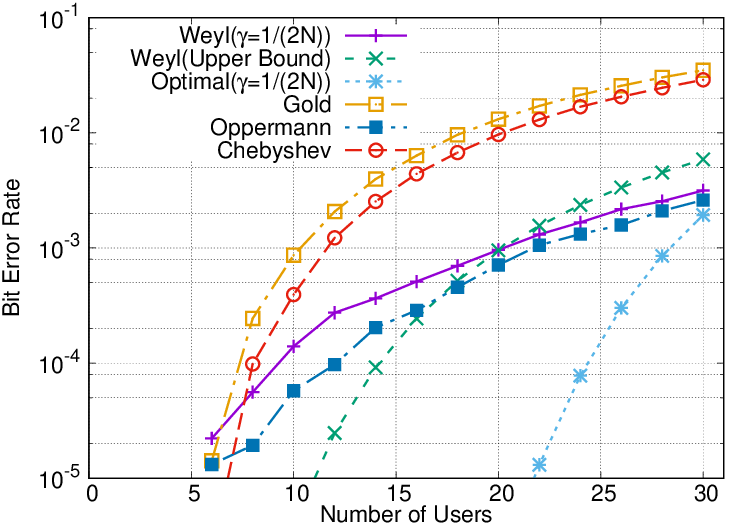} 
   \caption{Comparison with other sequences in BPSK systems: relation \newline between the number of users and BER: $E/N_0 = 25$(db), $N$=31}
   \label{fig:ber_user}
\end{minipage}
\begin{minipage}{0.50\hsize}
   \centering
   \includegraphics[width=3in]{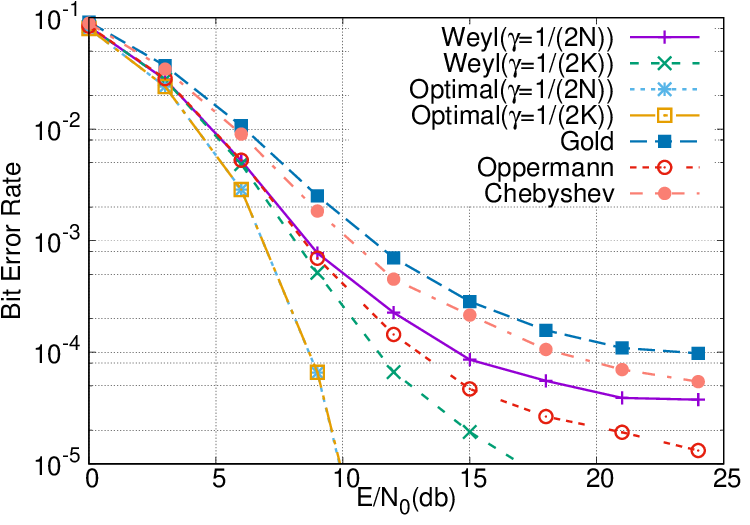} 
   \hspace{5mm}
   \caption{Comparison with other sequences in BPSK systems: relation \newline between the $E/N_0$(db) and BER: $K=7$, $N$=31}
   \label{fig:db}
\end{minipage}\\

\begin{minipage}{0.50\hsize}
   \centering
   \includegraphics[width=3in]{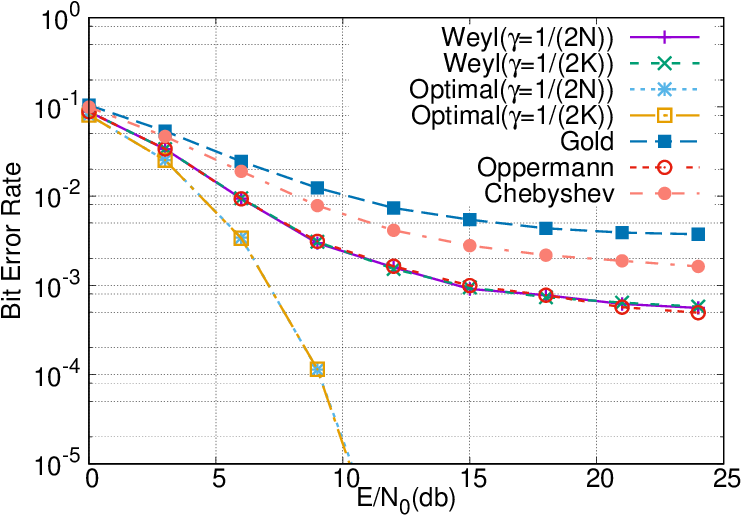} 
   \hspace{5mm}
   \caption{Comparison with other sequences in QPSK systems: relation \newline between the $E/N_0$(db) and BER: $K=7$, $N$=31}
   \label{fig:db_qpsk}
\end{minipage}
\begin{minipage}{0.50\hsize}
   \centering
   \includegraphics[width=3in]{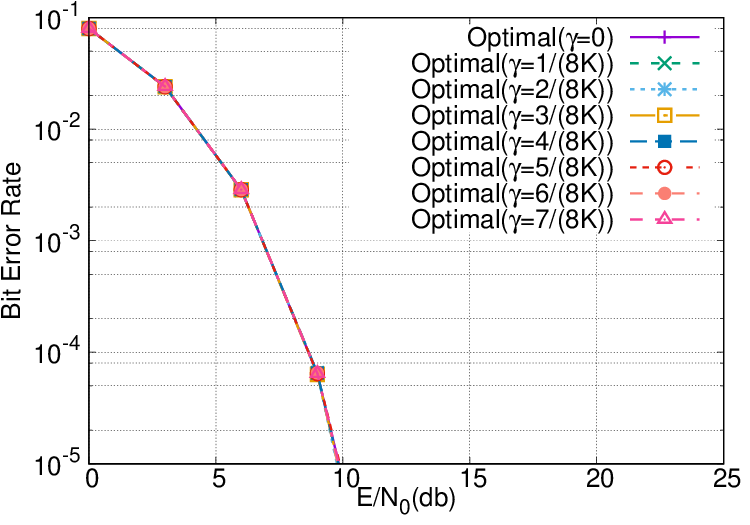} 
   \hspace{5mm}
   \caption{BER of optimal sequences with various $\gamma$ in BPSK \newline systems: relation between the $E/N_0$(db) and BER: $K=7$, $N$=31}
   \label{fig:db_opt}
   \end{minipage}
\end{figure*}

Figure \ref{fig:ber_user} shows the relation between the number of users and BER when $N=31$ and $E/N_0 = 25$(db), where $E$ is the energy per data bit. In this figure, we set $\gamma=\frac{1}{2N}$. The BER of the Weyl spreading sequences is lower than the one of the Gold codes and the optimal Chebyshev spreading sequences. However, it is higher than one of the Oppermann sequences. The upper bound is established when the number of the users $K$ is larger than 20. On the other hands, the BER of the global solution of the problem $(P)$, the Optimal Weyl sequences $\left\{\tilde{w}_{k,n}(K,\gamma)\right\}$ have the lowest BER. These sequences are dramatically efficient when the number of the users $K$ is fixed.

Figure \ref{fig:db} shows the relation between the $E/N_0$(db) and BER in $K = 7$.  
In this figure, the BER of the Weyl spreading sequences $\left\{\tilde{w}_{k,n}(N,\frac{1}{2K})\right\}$ is lower than one of the Oppermann sequences. This result shows that BER of the Weyl spreading sequences is changed when the value of $\gamma$ is varied. 

As a modulation technique, a Quadrature Phase-Shift Keying (QPSK) modulation is often used. In AWGN channels, the relation of BER between BPSK and QPSK systems is known with a given Signal to Noise Ratio (SNR) \cite{ofdmcdma}. Figure \ref{fig:db_qpsk} shows the relation between the $E/N_0$(db) and BER in QPSK systems and $K=7$. Similar to BPSK systems, the optimal sequences have the lowest BER in QPSK systems and the BER of the optimal sequences is independent of the parameter $\gamma$. Further, the BER of Weyl spreading sequences does not depend on the parameter $\gamma$ while BER depends on $\gamma$ in BPSK systems. This result implies that the distribution of the inter-symbol interference of the Weyl spreading sequences is not Gaussian. The reason for this is stated as follows. In BPSK and QPSK systems, the BER is known when the SINR is given and inter-symbol interference obeys Gaussian. As seen in Fig. \ref{fig:db}, with the Weyl spreading sequences, the BER curves depend on $\gamma$ in BPSK systems. Therefore, if the distribution of the inter-symbol interference of the Weyl spreading sequences is Gaussian, then the BER curves should depend on the parameter $\gamma$ in QPSK systems. However, as seen in Fig \ref{fig:db_qpsk}, the BER curves are independent of the parameter $\gamma$ in QPSK systems. Thus, we conclude that the distribution of inter-symbol interference is not Gaussian.

In Section 3, we have shown that the optimal sequences have an arbitrary parameter $\gamma$. It is unknown whether the BER of the optimal sequences is independent of $\gamma$ or not since we minimize the upper bound of inter-symbol interference. We numerically verify whether the BER of the optimal sequences is independent of $\gamma$ or not. We choose the parameters $N=31$ and $K=7$ and a system is a BPSK system. Figure \ref{fig:db_opt} shows the BER curves in various parameters $\gamma=\{0,1/(8K),\ldots,7/(8K)\}$. As seen in Fig \ref{fig:db_opt}, each BER of the optimal sequences is the same. Therefore, we conclude that the BER of the optimal sequences is independent of the parameter $\gamma$.

\subsection{Comparison with Systematic Approaches}
\begin{figure*}[htbp] 
\begin{minipage}{0.50\hsize}
   \centering
   \includegraphics[width=3in]{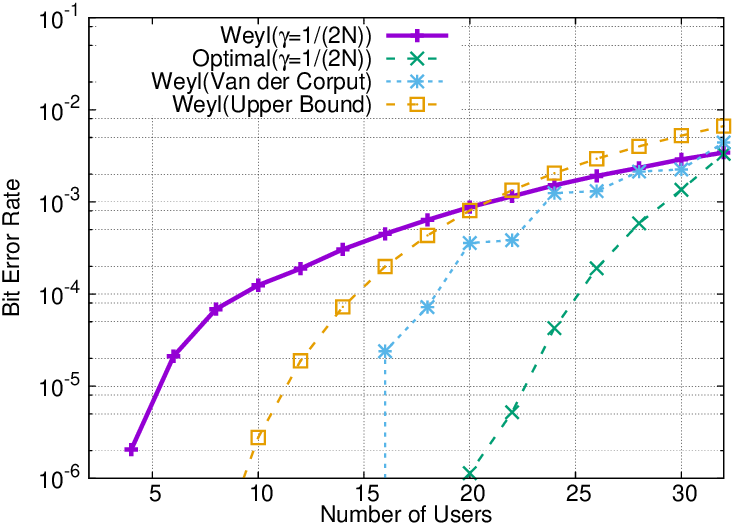} 
   \caption{Comparison with systematic approaches in BPSK systems: relation \newline between the number of the users and BER: $N=32$}
   \label{fig:sys}
\end{minipage}
\begin{minipage}{0.50\hsize}
   \centering
   \includegraphics[width=3in]{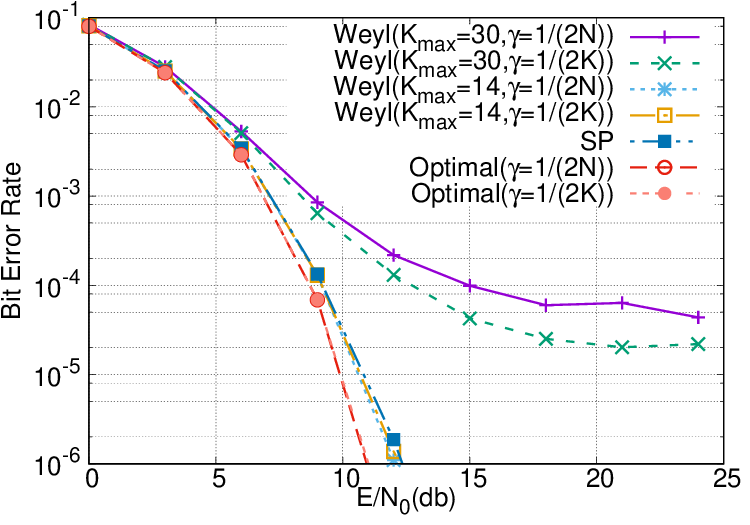} 
   \caption{Comparison with SP sequence in BPSK systems: relation \newline between the $E/N_0$(db) and BER: $K=7$, $N=30$}
   \label{fig:sp}
   \end{minipage}
\end{figure*}

In Section 5, we have discussed how to assign the element $\sigma_k$ to the user $k$ and proposed the method to assign. We set the length $N=32$ and the parameter $\gamma=\frac{1}{2N}$. We compare two types of  Weyl spreading sequences, whose $\sigma_k$ is randomly assigned to user $k$ and whose $\frac{\sigma_k}{N}$ is orderly assigned as the $k$-th element of the Van der Corput sequence. Note that the first $K$ elements of the Van der Corput sequence are used as $\frac{\sigma_k}{N}$ when the number of the users is $K$. Figure \ref{fig:sys} shows the relation between the number of users and BER when $E/N_0 = 25$(db). The BER of the spreading sequences obtained by Eq. (\ref{eq:vander_seq}) is lower than one of the spreading sequences to which we randomly assign $\sigma_k$. The BER curve of sequences obtained by Eq. (\ref{eq:vander_seq}) is not smooth. This result is caused by systematic assignment of $\sigma_k$. From Fig. \ref{fig:sys}, we conclude that the Weyl spreading sequences will have better performance if we successfully assign $\sigma_k$.

\subsection{Comparison with SP Sequence}
The Song-Park (SP) sequences have been proposed in \cite{song_park}. We set the length $N=30$ and the number of the users $K=7$. Then, the maximum number of the users of the SP sequences is 14. Thus, we compare them with four types of the Weyl spreading sequences. We choose the parameters as $(K_{\max},\gamma) = \{(30,1/(2N)), (30,1/(2K)), (14,1/(2N)), (14, 1/(2K))\}$. The parameter $K_{\max} = 14$ represents the situation where the maximum number of the users is 14. Thus, the Weyl spreading sequences $\left\{\tilde{w}_{k,n}(14,\gamma)\right\}$ have the same feature to the SP sequences. Figure \ref{fig:sp} shows the relation between the $E/N_0$(db) and BER. The BER of the Weyl spreading sequences whose $K_{\max} = 30$ is higher than one of the SP sequences. However, the BER of the Weyl spreading sequences whose $K_{\max} = 14$ and one of the SP sequences are the same. Further, the BER of the Weyl spreading sequences whose $K_{\max} = 14$ is the same BER in different $\gamma$. These results could suggest that the optimal parameter $\gamma$ will depend on $N$, $K$ and $K_{\max}$. The BER of the global solutions is lowest and each BER of them is the same in $\gamma = \frac{1}{2N}$ and $\frac{1}{2K}$. This result corresponds to our conclusion that the BER of the optimal sequences is independent of $\gamma$. 

\section{Conclusion}
In this paper, we have defined the Weyl sequence class and shown the features of the sequences in the class. We have constructed the optimization problem: minimize the upper bound of the absolute value of the whole inter-symbol interference and derived the global solutions. From this solution, we can derive other sequences, the Sarwate's sequences, and the SP sequences. We also have evaluated their SINR in a special case and shown the simulation results for an asynchronous CDMA system. From these results, the global solution is significantly efficient when the number of the users $K$ is fixed. Moreover, the performance of the global solution is independent of the parameter $\gamma$.

In the global solution of the problem $(P)$, the parameter $\gamma$ is any real number. However, its BER depends on $\gamma$ when we let the maximum number of users $K_{\max}$ be $N$ or any number other than $K$. The remained issue is to investigate the optimal $\gamma$ and how to assign $\sigma_k$ successfully to the user $k$.

%

\appendix
\section{}

In this appendix, we prove that the global optimal solutions of $(P)$, $\rho^*_i$ and $t^*_{i,k}$ are given by
\begin{equation}
\begin{split}
\rho^*_i &= \gamma + \frac{i-1}{K} \hspace{2mm}(i=1,2,\ldots,K),\\
t^*_{i,k} &= \min\left\{\frac{|k - i|}{K},1-\frac{|k - i|}{K}\right\},
\end{split}
\label{eq:gs}
\end{equation}
where $\gamma$ is a real number. \\

Since the problem $(P)$ is a convex, it is necessary and sufficient for the global solution to satisfy the KKT conditions, Eqs. (\ref{eq:KKT1_1})-(\ref{eq:KKT1_3}).\\
From Eq. (\ref{eq:KKT1_3}), when $\rho^*_i$ satisfies Eq. (\ref{eq:gs}), it is clearly that
\begin{equation}
\begin{split}
\nu_{i} &= 0 \hspace{2mm}(i = 1,2,\ldots,K-1),\\
o_{i,k} &= 0 \hspace{2mm}(i < k)
\end{split}
 \end{equation}
 since $e_i(\mathbf{x}^*) < 0$ and $h_{i,k}(\mathbf{x}^*) < 0$.
 We let $\xi_{1} = \xi_{K} = 0$. Thus, it is sufficient to consider only two kinds of the Lagrange multipliers, $\lambda_{i,k}$ and $\mu_{i,k}$. They satisfy the following equation which is obtained from Eq. (\ref{eq:KKT1_1}):
 \begin{equation}
  \begin{split}
 &-\sum_{i<k}\frac{\pi \cos(\pi t^*_{i,k})}{\sin^2(\pi t^*_{i,k})}
 \left( \begin{array}{c}
 \mathbf{0}\\
 \mathbf{e}_{i,k}
\end{array}
\right) + 
\sum_{i<k}\lambda_{i,k}
 \left( \begin{array}{c}
 \mathbf{e}_i - \mathbf{e}_k\\
 \mathbf{e}_{i,k}
\end{array}
\right) \\
+&\sum_{i<k}\mu_{i,k}
 \left( \begin{array}{c}
 -\mathbf{e}_i + \mathbf{e}_k\\
 \mathbf{e}_{i,k}
\end{array}
\right) = \mathbf{0},
 \end{split}
\label{eq:KKT2}
 \end{equation}
 where $\mathbf{e}_i \in \mathbb{R}^K$ have $1$ in the $i$-th element and 0 in the others and $\mathbf{e}_{i,k} \in \mathbb{R}^{K(K-1)/2}$ have 
 $1$ in the $\displaystyle \left\{i(2K -i - 1)/2 + k - K\right\}$-th element and 0 in the others. From Eq. (\ref{eq:KKT2}), we consider two vector equations. One is the first $K$-dimensional vector equation of Eq. (\ref{eq:KKT2}) and the other is the last $K(K-1)/2$-dimensional vector equation. They are expressed as
 \begin{equation}
 \sum_{i<k}(\lambda_{i,k} - \mu_{i,k})(\mathbf{e}_i - \mathbf{e}_k) = \mathbf{0}\label{eq:sa1},
  \end{equation}
   \begin{equation}
 \sum_{i<k}\left(\frac{\pi \cos(\pi t^*_{i,k})}{\sin^2(\pi t^*_{i,k})} -\lambda_{i,k}- \mu_{i,k}\right)\mathbf{e}_{i,k}=\mathbf{0}.\label{eq:sa2}
 \end{equation}
Then, we define $\alpha(t^*_{i,k})$ as
 \begin{equation}
 \alpha(t^*_{i,k}) = \frac{\pi \cos(\pi t^*_{i,k})}{\sin^2(\pi t^*_{i,k})}.
 \end{equation}
 Note that $\alpha(t^*_{i,k}) \geq 0$ since $0 < t^*_{i,k} \leq \frac{1}{2}$. From the definition of $t^*_{i,k}$, $\alpha(t^*_{i,k})$ only depends on the absolute value of difference, $|k-i|$. We therefore rewrite $\alpha(t^*_{i,k})$ as
 \begin{equation}
 \alpha(t^*_{i,k}) = \tilde{\alpha}(|k-i|).
 \end{equation}
 The variable $\tilde{\alpha}(|k-i|)$ has the property that
  \begin{equation}
  \tilde{\alpha}(k)=\tilde{\alpha}(K - k)\hspace{2mm}(1 \leq k \leq K)\label{eq:alpha_pro}.
  \end{equation}
 This result is obtained from the definition of $t^*_{i,k}$. We consider the two types of $K$: $K$ is an odd number or $K$ is an even number.
\subsection{$K$ is an odd number}
 For all $i$ and $k\hspace{2mm}(i<k)$, $\rho^*_i$, $\rho^*_k$ and $t^*_{i,k}$ satisfy either only $c_{i,k}(\mathbf{x^*}) = 0$ or $d_{i,k}(\mathbf{x^*}) = 0$. They satisfy
\begin{equation}
\begin{split}
 c_{i,k}(\mathbf{x}^*) &= 0, d_{i,k}(\mathbf{x}^*) < 0, (k-i < K/2),\\ 
 d_{i,k}(\mathbf{x}^*) &= 0, c_{i,k}(\mathbf{x}^*) < 0,  (k-i > K/2),\\
 \lambda_{i,k} &= \left\{ \begin{array}{c c}
\tilde{\alpha}(k-i)& (k-i < K/2)\\
 0 & (k-i > K/2)
 \end{array}
 \right. ,\\
 \mu_{i,k} &= \left\{ \begin{array}{c c}
0 & (k-i < K/2)\\
\tilde{\alpha}(k-i) & (k-i > K/2)
 \end{array}
 \right. .
 \end{split}
  \label{eq:con1}
 \end{equation}
 We consider the $n$-th element of the left side of Eq. (\ref{eq:sa1}). 
 \mathindent=0mm
 \begin{equation}
 \begin{split}
 & \sum_{n<k}(\lambda_{n,k} - \mu_{n,k}) - \sum_{i<n}(\lambda_{i,n} - \mu_{i,n})\\
 =& \sum_{\substack{n<k \\ k-n<K/2}}\lambda_{n,k} + \sum_{\substack{i<n \\ n-i>K/2}}\mu_{i,n} -\sum_{\substack{i<n \\ n-i<K/2}}\lambda_{i,n} - \sum_{\substack{n<k \\ k-n>K/2}}\mu_{n,k}\\
 =& \sum_{\substack{n<k \\ k-n<K/2}}\tilde{\alpha}(k-n) + \sum_{\substack{i<n \\ n-i>K/2}}\tilde{\alpha}(n-i) \\
 & -\sum_{\substack{i<n \\ n-i<K/2}}\tilde{\alpha}(n-i) - \sum_{\substack{n<k \\ k-n>K/2}}\tilde{\alpha}(k-n) \\
 =& \sum_{\substack{n<k \\ k-n<K/2}}\tilde{\alpha}(k-n) + \sum_{\substack{i<n \\ n-i>K/2}}\tilde{\alpha}(K+i-n) \\
 & -\sum_{\substack{i<n \\ n-i<K/2}}\tilde{\alpha}(n-i)  -\sum_{\substack{n<k \\ k-n>K/2}}\tilde{\alpha}(K+n-k) \\
 =& \sum_{\substack{n<k \\ k-n<K/2}}\tilde{\alpha}(k-n) + \sum_{\substack{i<n \\ n-i<K/2}}\tilde{\alpha}(n-i)\\
  & -\sum_{\substack{i<n \\ n-i<K/2}}\tilde{\alpha}(n-i)   -\sum_{\substack{n<k \\ k-n<K/2}}\tilde{\alpha}(k-n)=0. 
 \end{split}
    \end{equation}
    \mathindent=7mm
  From Eq. (\ref{eq:con1}), for all the integers $i$ and $k$, the term in summation of the left side of Eq. (\ref{eq:sa2}) equals $0$.
 From the above proof, all the Lagrange multipliers satisfy Eq. $(\ref{eq:KKT1_3})$.
\subsection{$K$ is an even number}
The Lagrange multipliers $\rho^*_i$, $\rho^*_k$ and $t^*_{i,k}$ satisfy
 \begin{equation}
 \begin{split}
 c_{i,k}(\mathbf{x^*}) &= 0, d_{i,k}(\mathbf{x^*}) < 0, \hspace{3mm}(k-i < K/2),\\ 
 d_{i,k}(\mathbf{x^*}) &= 0, c_{i,k}(\mathbf{x^*}) < 0,  \hspace{3mm}(k-i > K/2),\\
  d_{i,k}(\mathbf{x^*}) &= 0, c_{i,k}(\mathbf{x^*}) = 0, \hspace{3mm} (k-i = K/2).
  \end{split}
  \end{equation}
  When $k-i = K/2$, they satisfy $ c_{i,k}(\mathbf{x^*}) = 0$ and $d_{i,k}(\mathbf{x^*}) = 0$. Thus, we set
  \begin{equation}
  \begin{split}
 \lambda_{i,k} &= \left\{ \begin{array}{c c}
 \tilde{\alpha}(k-i) & (k-i < K/2),\\
  \tilde{\alpha}(k-i)/2 & (k-i = K/2),\\
 0 & (k-i > K/2)
 \end{array}
 \right. \\
 \mu_{i,k} &= \left\{ \begin{array}{c c}
0 & (k-i < K/2),\\
  \tilde{\alpha}(k-i)/2 & (k-i = K/2),\\
 \tilde{\alpha}(k-i) & (k-i > K/2)
 \end{array}
 \right. 
   \end{split}
 \end{equation}
 Similar to the case that $K$ is an odd number, we consider the $n$-th element of left side of Eq. (\ref{eq:sa1}). 
 \mathindent=0mm
 \begin{equation}
 \begin{split}
 &\sum_{n<k}(\lambda_{n,k} - \mu_{n,k})-\sum_{i<n}(\lambda_{i,n} - \mu_{i,n}) \\
 =& \sum_{\substack{n<k \\ k-n<K/2}}\lambda_{n,k}  - \sum_{\substack{n<k \\ k-n>K/2}}\mu_{n,k} - \sum_{\substack{i<n \\ n-i<K/2}}\lambda_{i,n} + \sum_{\substack{i<n \\ n-i>K/2}}\mu_{i,n}  \\
 &+\sum_{\substack{n<k \\ k-n=K/2}}\lambda_{n,k}  - \sum_{\substack{n<k \\ k-n=K/2}}\mu_{n,k} -\sum_{\substack{i<n \\ n-i=K/2}}\lambda_{i,n} + \sum_{\substack{i<n \\ n-i=K/2}}\mu_{i,n} .
  \end{split}
 \end{equation}
 \mathindent=7mm
 The terms of the difference equaling $K/2$ vanish. Therefore, we obtain
 \mathindent=0mm
  \begin{equation}
  \begin{split}
   & \sum_{n<k}(\lambda_{n,k} - \mu_{n,k})-\sum_{i<n}(\lambda_{i,n} - \mu_{i,n}) \\
   =&\sum_{\substack{n<k \\ k-n<K/2}}\lambda_{n,k}  - \sum_{\substack{n<k \\ k-n>K/2}}\mu_{n,k} - \sum_{\substack{i<n \\ n-i<K/2}}\lambda_{i,n} + \sum_{\substack{i<n \\ n-i>K/2}}\mu_{i,n}\\
  =& \sum_{\substack{n<k \\ k-n<K/2}} \tilde{\alpha}(k-n) - \sum_{\substack{n<k \\ k-n>K/2}}  \tilde{\alpha}(k-n)\\
 &-\sum_{\substack{i<n \\ n-i<K/2}}  \tilde{\alpha}(n-i) + \sum_{\substack{i<n \\ n-i>K/2}}  \tilde{\alpha}(n-i)\\
  =& \sum_{\substack{n<k \\ k-n<K/2}} \tilde{\alpha}(k-n) - \sum_{\substack{n<k \\ k-n>K/2}}  \tilde{\alpha}(K-k+n)\\
 &-\sum_{\substack{i<n \\ n-i<K/2}}  \tilde{\alpha}(n-i) + \sum_{\substack{i<n \\ n-i>K/2}}  \tilde{\alpha}(K-n+i)\\
  =& \sum_{\substack{n<k \\ k-n<K/2}} \tilde{\alpha}(k-n) - \sum_{\substack{n<k \\ k-n<K/2}}  \tilde{\alpha}(k-n)\\
 &-\sum_{\substack{i<n \\ n-i<K/2}}  \tilde{\alpha}(n-i) + \sum_{\substack{i<n \\ n-i<K/2}}  \tilde{\alpha}(n-i)=0.
 \end{split}
 \label{eq:LA2}
   \end{equation}
   \mathindent=7mm
   Thus, we have proven that Eq. (\ref{eq:LA2}) equals to 0. It is clearly that the left side of Eq. (\ref{eq:sa2}) equals $0$ when $k-i \neq K/2$. When $k-i = K/2$, it follows that
   \begin{equation}
   \tilde{\alpha}(K/2) - \frac{\tilde{\alpha}(K/2)}{2} - \frac{\tilde{\alpha}(K/2)}{2} = 0.
   \end{equation}
   Thus, for all the integer $i$ and $k$, Eq. (\ref{eq:sa2}) is satisfied.\\
   
  From the proofs A and B, we have proven that the existence of the Lagrange multipliers which satisfy Eq. $(\ref{eq:KKT1_3})$.
 Therefore, $\rho_i^*$ and $t^*_{i,k}$ are the global solutions of the problem ($P$).
 
\section{}
In this appendix, with the spreading sequences $\{\tilde{w}_{k,n}(N,\gamma)\}$, we prove that SINR of the user $i$ is given by
\begin{equation}
\mbox{SINR}_i=\left\{ R_i +  \frac{N_0}{2E} \right\}^{-1/2},
\end{equation}
where
\mathindent=0mm
\begin{equation}
R_i = \frac{(K-1)}{18N^2}\left\{2(N+1) +\left(N-2\right)\cos\left(2 \pi \left(\gamma + \frac{\sigma_i}{N}\right)\right)\right\}.
\end{equation}
\mathindent=7mm
We assume that the element $\sigma_k$ is a random variable uniformly distributed in $\{0,1,2,\ldots,N-1\}$. This assumption is fulfilled when the ratio $K/N$ is close to $1$, that is, the number of users is sufficiently large since SINR is not the reciprocal of the average of the inter-symbol interference over the users. However, with the spreading sequences $\{\tilde{w}_{k,n}(N,\gamma)\}$, they are the same when the number of users $K$ equals $N$ (see Eq. (\ref{eq:first_r2})). Thus, it is conceivable that the assumption is established when the ratio $K/N$ is close to $1$.

The correlation function $C_{i,k}(l)$ of the spreading sequences in Eq. (\ref{eq:optimal_seq}) is 
\begin{equation}
C_{i,k}(l) = \left\{ 
\begin{array}{c l}
\displaystyle -Z_{\sigma_i,\sigma_k}\Phi_{\gamma,\sigma_i,\sigma_k}(l) & 0 \leq l \leq N-1,\\
\displaystyle Z_{\sigma_i,\sigma_k}\Phi_{\gamma,\sigma_i,\sigma_k}(l) & 1-N \leq l < 0,\\
0 & | l | \geq N,\\
\end{array} \right. ,
\end{equation}
where 
\begin{equation}
 Z_{\sigma_i,\sigma_k} = \frac{\exp\left(2 \pi j \frac{\sigma_k - \sigma_i}{N}\right)}{1 - \exp\left(2 \pi j \frac{\sigma_k - \sigma_i}{N}\right)}
 \end{equation}
and
\mathindent=0mm
\begin{equation}
\Phi_{\gamma,\sigma_i,\sigma_k}(l) =  \exp\left(-2\pi j l\left(\gamma + \frac{\sigma_k}{N}\right)\right) - \exp\left(-2\pi j l\left(\gamma + \frac{\sigma_i}{N}\right)\right).
\end{equation}
\mathindent=7mm
Thus, we obtain the squared absolute value of $C_{i,k}(l)$:
\begin{equation}
|C_{i,k}(l)| ^2 = \frac{1 -  \cos\left(2\pi l \frac{\sigma_k - \sigma_i}{N}\right)}{1 - \cos\left(2 \pi \frac{\sigma_k - \sigma_i}{N}\right)}.
\end{equation}
On the other hand, the following relations are satisfied:
\mathindent=0mm
\begin{equation}
\begin{split}
&\sum_{l=0}^{N-1}|C_{i,k}(l - N)|^2  = \sum_{l=0}^{N-1}|C_{i,k}\left(l - N + 1)\right|^2  = \sum_{l=0}^{N-1}|C_{i,k}\left(l\right)|^2\\
 =& \sum_{l=0}^{N-1}|C_{i,k}\left(l+1\right)|^2 =  \frac{N}{1 - \cos\left(2 \pi \frac{\sigma_k - \sigma_i}{N}\right)}, 
  \end{split}
 \label{eq:c_1}
\end{equation}
\mathindent=7mm
\begin{equation}
\begin{split}
&\sum_{l=0}^{N-1}\operatorname{Re}[C_{i,k}\left(l-N\right)\overline{C_{i,k}\left(l-N+1\right)}]\\
=&\frac{N\left\{\cos\left(2 \pi \left(\gamma + \frac{\sigma_k}{N}\right)\right) +\cos\left(2 \pi \left(\gamma + \frac{\sigma_i}{N}\right)\right)\right\}}{2\left(1-\cos\left(2 \pi \frac{\sigma_k - \sigma_i}{N}\right)\right)},
 \end{split}
\end{equation}
and
\begin{equation}
\begin{split}
&\sum_{l=0}^{N-1}\operatorname{Re}\left[C_{i,k}\left(l-N\right)\overline{C_{i,k}\left(l-N+1\right)}\right]\\
 =& \sum_{l=0}^{N-1}\operatorname{Re}\left[C_{i,k}\left(l\right)\overline{C_{i,k}\left(l+1\right)}\right].
 \label{eq:c_2}
 \end{split}
\end{equation}
In the above equations, we used the assumption $\sigma_i \neq \sigma_k$.
From Eqs. (\ref{eq:c_1})-(\ref{eq:c_2}), $r_{i,k}$ in Eq. (\ref{eq:SNR}) is given by
\begin{equation}
\begin{split}
r_{i,k} =& \frac{N}{1 - \cos\left(2 \pi \frac{\sigma_k - \sigma_i}{N}\right)}\\
&\cdot\left\{4 + \cos\left(2 \pi \left(\gamma + \frac{\sigma_k}{N}\right)\right) +\cos\left(2 \pi \left(\gamma + \frac{\sigma_i}{N}\right)\right)\right\}.
\label{eq:r_ik}
 \end{split}
\end{equation}
When we calculate the sum of Eq. (\ref{eq:r_ik}), the first term of it is given by
\begin{equation}
\sum_{\substack{ k=1\\ k \neq i}} \frac{4N}{1 - \cos\left(2 \pi \frac{\sigma_k - \sigma_i}{N}\right)} = \sum_{\substack{ k=1\\ k \neq i}} \frac{2N}{\sin^2\left(\pi \frac{\sigma_k - \sigma_i}{N}\right)}.
\label{eq:first_r}
\end{equation}
The integer $\sigma_k \in \{0,1,2,\ldots,N-1\}$ is a random variable and satisfies $\sigma_k \neq \sigma_i$ when $k \neq i$. Thus, $\sigma_k$ is expressed as
\begin{equation}
\sigma_k = (\sigma_i + q) \bmod N, \hspace{2mm} q \in \{1,2,\ldots,N-1\}.
\end{equation}
and we can treat $q$ as a random variable instead of $\sigma_k$. The integer $q$ is uniformly distributed in $\{1,2,\ldots,N-1\}$. Thus, the average of Eq. (\ref{eq:first_r}) is
\begin{equation}
\begin{split}
&\operatorname{E}\left\{\sum_{\substack{ k=1\\ k \neq i}} \frac{2N}{\sin^2\left(\pi \frac{\sigma_k - \sigma_i}{N}\right)} \right\}\\
 =& \sum_{\substack{ k=1\\ k \neq i}} \operatorname{E}\left\{\frac{2N}{\sin^2\left(\pi \frac{\sigma_k - \sigma_i}{N}\right)} \right\}\\
=& \sum_{\substack{ k=1\\ k \neq i}} \frac{1}{N-1}\sum_{q=1}^{N-1}\frac{2N}{\sin^2(\pi \frac{q}{N})}\\
=&\frac{K-1}{N-1}\sum_{q=1}^{N-1}\frac{2N}{\sin^2\left(\pi \frac{q}{N}\right)},
\end{split}
\label{eq:first_r2}
\end{equation}
where $\operatorname{E}$ is the average over $\sigma_k$.

In \cite{hansen}, it is shown that
\begin{equation}
\sum^{n-1}_{k=1}\frac{1}{\sin^2(\pi\frac{k}{n})}=\frac{n^2-1}{3}=\frac{(n-1)(n+1)}{3}.
\end{equation}
Thus, Eq. (\ref{eq:first_r2}) is equivalent to
\begin{equation}
\frac{K-1}{N-1}\sum_{q=1}^{N-1}\frac{2N}{\sin^2\left(\pi \frac{q}{N}\right)}=\frac{2N\left(N+1\right)\left(K-1\right)}{3}.
\end{equation}
From the above result, we obtain the following relation 
\begin{equation}
\operatorname{E}\left\{\sum_{\substack{ k=1\\ k \neq i}} \frac{4N}{ 1 - \cos\left(2 \pi \frac{\sigma_k - \sigma_i}{N}\right)}\right\} = \frac{2N\left(N+1\right)\left(K-1\right)}{3}. 
\label{eq:r1}
\end{equation}
The average of the second term of the sum of Eq. (\ref{eq:r_ik}) is given by
\begin{equation}
\begin{split}
& \sum_{\substack{ k=1\\ k \neq i}}\operatorname{E}\left\{\frac{N\cos\left(2 \pi \left(\gamma + \frac{\sigma_k}{N}\right)\right)}{1 - \cos\left(2 \pi \frac{\sigma_k - \sigma_i}{N}\right)}\right\}\\
=&\frac{K-1}{N-1}\sum_{q=1}^{N-1}\frac{N}{2\sin^2\left(\pi \frac{q}{N}\right)}\cos\left(2 \pi \left(\gamma + \frac{\sigma_i+q}{N}\right)\right)\\
=& \frac{N(K-1)}{N-1}\sum_{q=1}^{N-1}\left\{\frac{\cos\left(2 \pi \left(\gamma + \frac{\sigma_i}{N}\right)\right)\cos\left(2 \pi \frac{q}{N}\right)}{2\sin^2\left(\pi \frac{q}{N}\right)}\right. \\
&-\left.\frac{\sin\left(2 \pi \left(\gamma + \frac{\sigma_i}{N}\right)\right)\sin\left(2 \pi \frac{q}{N}\right)}{2\sin^2\left(\pi \frac{q}{N}\right)}\right\} \\
=& \frac{N(K-1)}{N-1}\sum_{q=1}^{N-1}\left\{\frac{\cos\left(2 \pi \left(\gamma + \frac{\sigma_i}{N}\right)\right)\left(1-2\sin^2\left(\pi \frac{q}{N}\right)\right)}{2\sin^2\left(\pi \frac{q}{N}\right)}\right. \\
&-\left.\frac{\sin\left(2 \pi \left(\gamma + \frac{\sigma_i}{N}\right)\right)\cos\left(\pi \frac{q}{N}\right)}{\sin\left(\pi \frac{q}{N}\right)}\right\}.
\end{split}
\label{eq:term2}
\end{equation}
Note that it is clear that
\begin{equation}
\sum_{q=1}^{N-1} \frac{\cos\left(\pi \frac{q}{N}\right)}{\sin\left(\pi \frac{q}{N}\right)} = 0.
\end{equation} 
 Therefore, Eq. (\ref{eq:term2}) is rewritten as
\begin{equation}
\begin{split}
 &\frac{N(K-1)}{N-1}\cos\left(2 \pi \left(\gamma + \frac{\sigma_i}{N}\right)\right)\sum_{q=1}^{N-1}\left\{\frac{1}{2\sin^2\left(\pi \frac{q}{N}\right)} -1 \right\}\\
 =&N(K-1)\left(\frac{N+1}{6} - 1\right) \cos\left(2 \pi \left(\gamma + \frac{\sigma_i}{N}\right)\right)\\
 =&\frac{N(N-5)(K-1)}{6}\cos\left(2 \pi \left(\gamma + \frac{\sigma_i}{N}\right)\right).
\end{split}
\end{equation}
Thus, the sum of the average of the second term in Eq. (\ref{eq:r_ik}) is written as
\begin{equation}
\begin{split}
& \sum_{\substack{k=1 \\ k \neq i}}^K\operatorname{E}\left\{\frac{N\cos\left(2 \pi \left(\gamma + \frac{\sigma_k}{N}\right)\right)}{1 - \cos\left(2 \pi \frac{\sigma_k - \sigma_i}{N}\right)}\right\}\\
=&\frac{N(N-5)(K-1)}{6}\cos\left(2 \pi \left(\gamma + \frac{\sigma_i}{N}\right)\right)
\end{split}
\end{equation}
Similarly, we obtain the average of the sum of the third term of Eq. (\ref{eq:r_ik}):
\begin{equation}
\begin{split}
&\operatorname{E}\left\{ \sum_{\substack{ k=1\\ k \neq i} }\frac{N\cos\left(2 \pi \left(\gamma + \frac{\sigma_i}{N}\right)\right)}{1-\cos\left(2 \pi \frac{\sigma_k - \sigma_i}{N}\right)} \right\} \\
=&\frac{K-1}{N-1}\sum_{q=1}^{N-1}\frac{N\cos\left(2 \pi \left(\gamma + \frac{\sigma_i}{N}\right)\right)}{2\sin^2\left(\pi \frac{q}{N}\right)}\\
=&\frac{N(N+1)(K-1)}{6}\cos\left(2 \pi \left(\gamma + \frac{\sigma_i}{N}\right)\right).
\end{split}
\label{eq:r2_1}
\end{equation}
Finally, we obtain the following relation
\begin{equation}
\begin{split}
&\operatorname{E}\left\{\sum_{\substack{ k=1\\ k \neq i} }\frac{N\left(\cos\left(2 \pi \left(\gamma + \frac{\sigma_k}{N}\right)\right) +\cos\left(2 \pi \left(\gamma + \frac{\sigma_i}{N}\right)\right)\right)}{1-\cos\left(2 \pi \frac{\sigma_k - \sigma_i}{N}\right)} \right\}\\
=&\frac{N(N-2)(K-1)}{3}\cos\left(2 \pi \left(\gamma + \frac{\sigma_i}{N}\right)\right).
\end{split}
\label{eq:r2_3}
\end{equation}
From Eq. (\ref{eq:r1}) and Eq. (\ref{eq:r2_3}), we arrive at SINR of the user $i$ with the spreading sequence $(\tilde{w}_{k,n})$ 
\begin{equation}
\mbox{SINR}_i=\left\{ R_i +  \frac{N_0}{2E} \right\}^{-1/2},
\end{equation}
where
\begin{equation}
R_i = \frac{(K-1)}{18N^2}\left\{2(N+1) + \left(N-2\right)\cos\left(2 \pi \left(\gamma + \frac{\sigma_i}{N}\right)\right)\right\}.
\end{equation}

\section*{Acknowledgment}
One of the authors, Hirofumi Tsuda, would like to thank for advise of Prof. Nobuo Yamashita and Dr. Shin-itiro Goto.


\begin{thebibliography}{99}
\bibitem{shannon} C. E. Shannon, ``A mathematical theory of communication,'' {\it Bell System Technical Journal}, Volume 27, Issue 3, 379-423 (1948).
\bibitem{efficiency} S. Verd\'u and S. Shamai, ``Spectral efficiency of CDMA with random spreading.'' {\it IEEE Transactions on Information Theory} 45.2 (1999): 622-640.
\bibitem{dscdma} J. Proakis, ``Digital Communications. 1995'', McGraw-Hill, New York.
\bibitem{multiple} R. Steele and L. Hanzo, ``Mobile Radio Communications'', Second and Third Generation Cellular and WATM Systems: 2nd. IEEE Press-John Wiley, 1999.
\bibitem{mmse} M. Honig, U. Madhow and S. Verdu, ``Blind adaptive multiuser detection.'' {\it IEEE Transactions on Information Theory} 41.4 (1995): 944-960.
\bibitem{ica_cdma} T. Ristaniemi and J. Joutsensalo, ``Advanced ICA-based receivers for block fading DS-CDMA channels'', {\it Signal Processing}, 82.3 (2002): 417-431.
\bibitem{ica} A. Hyv\"{a}rinen, J. Karhunen, and E. Oja, ``Independent Component Analysis'', Vol. 46. John Wiley $\&$ Sons, 2004.
\bibitem{ml} S. Verdu, ``Minimum probability of error for asynchronous Gaussian multiple-access channels'', {\it IEEE Transactions on Information Theory}, 32.1 (1986): 85-96.
\bibitem{sync} P. Viswanath, V. Anantharam, and D. N. C. Tse. "Optimal sequences, power control, and user capacity of synchronous CDMA systems with linear MMSE multiuser receivers." {\it IEEE Transactions on Information Theory} 45.6 (1999): 1968-1983.
\bibitem{welch} L. R. Welch, ``Lower bounds on the maximum cross correlation of signals'', {\it IEEE Transactions on Information Theory}, 20.3 (1974): 397-399.
\bibitem{chipsync} S. Ulukus, and R. D. Yates. "User capacity of asynchronous CDMA systems with matched filter receivers and optimum signature sequences." {\it IEEE Transactions on Information Theory} 50.5 (2004): 903-909.
\bibitem{fundamental} L. Cottatellucci, R. R. Muller, and M. Debbah. "Asynchronous CDMA systems with random spreading--Part I: Fundamental limits." {\it IEEE Transactions on Information Theory} 56.4 (2010): 1477-1497.
\bibitem{gold}R. Gold, ``Optimal binary sequences for spread spectrum multiplexing (Corresp.).'' {\it IEEE Transactions on Information Theory},
 13.4 (1967): 619-621.
 \bibitem{prop} D. V. Sarwate, and M. B. Pursley. "Crosscorrelation properties of pseudorandom and related sequences." {\it Proceedings of the IEEE 68.5} (1980): 593-619.
 \bibitem{kasami} T. Kasami, ``Weight distribution formula for some class of cyclic codes.'' {\it Coordinated Science Laboratory Report}, no. R-285 (1966).
 \bibitem{chaos_cdma} G. Heidari-Bateni and C. D. McGillem, ``A chaotic direct-sequence spread-spectrum communication system.'' {\it IEEE Transactions on Communications}, 42.234 (1994): 1524-1527.
\bibitem{chaos_mod} K.S. Halle, C.W. Wu, M. Itoh and L.O. Chua, ``Spread spectrum communication through modulation of chaos.'' {\it International Journal of Bifurcation and Chaos}, 3.02 (1993): 469-477.
\bibitem{logistic} Y. Soobul, K. Chady and H. C.S. Rughooputh, ``Digital chaotic coding and modulation in CDMA.'' {\it Africon Conference in Africa, 2002. IEEE AFRICON. 6th}. Vol. 2. IEEE, 2002.
\bibitem{ergotic} C.C. Chen, K. Yao, K. Umeno and E. Biglieri, ``Design of spread-spectrum sequences using chaotic dynamical systems and ergodic theory.'' {\it IEEE Transactions on Circuits and Systems I: Fundamental Theory and Applications}, 48.9 (2001): 1110-1114.
\bibitem{period} K. Umeno and K. Kitayama. ``Spreading sequences using periodic orbits of chaos for CDMA.'' {\it Electronics Letters} 35.7 (1999): 545-546.
\bibitem{autocor} G. Mazzini, R. Rovatti and G. Setti, ``Interference minimisation by autocorrelation shaping in asynchronous DS-CDMA systems: chaos-based spreading is nearly optimal.'' {\it Electronics Letters}, 35.13 (1999): 1054-1055.
\bibitem{mazzini} G. Mazzini, G. Setti, and R. Rovatti, ``Chaotic complex spreading sequences for asynchronous DS-CDMA. I. System modeling and results.'' {\it IEEE Transactions on Circuits and Systems I: Fundamental Theory and Applications}, 44.10 (1997): 937-947.
\bibitem{limit} R. Riccardo, G. Mazzini and G. Setti, ``On the ultimate limits of chaos-based asynchronous DS-CDMA-I: basic definitions and results'', {\it IEEE Transactions on Circuits and Systems} I,  51.7 (2004): 1336-1347.
\bibitem{sarwate} D. V. Sarwate, ``Bounds on crosscorrelation and autocorrelation of sequences'', {\it IEEE Transactions on Information Theory}, 25.6 (1979): 720-724.
\bibitem{zadoff} R. Frank, S. Zadoff and R. Heimiller, ``Phase shift pulse codes with good periodic correlation properties (corresp.).'' {\it IRE Transactions on Information Theory} 8.6 (1962): 381-382.
\bibitem{chu}D. Chu, ``Polyphase codes with good periodic correlation properties (corresp.).'' {\it IEEE Transactions on Information Theory} 18.4 (1972): 531-532.
\bibitem{FZC_fam} J. Oppermann and B. S. Vucetic, ``Complex spreading sequences with a wide range of correlation properties.'' {\it IEEE Transactions on Communications}, 45.3 (1997): 365-375.
 \bibitem{weyl}H. Weyl. \"Uber die Gleichverteilung von Zahlen mod. Eins. Math. Ann, 77:313-352, 1916. (In German).
 \bibitem{dp} J. Dick and F. Pillichshammer, ``Digital nets and sequences: Discrepancy Theory and Quasi-Monte Carlo Integration''. Cambridge University Press, 2010.
 \bibitem{almost}K. Umeno, "Spread Spectrum Communications Based on Almost Periodic Functions " {\it IEICE Technical Report}, NLP 2014-101, pp. 11-16 (2014)(In Japanese)
 \bibitem{cross} D. V. Sarwate and M. B. Pursley, ``Crosscorrelation properties of pseudorandom and related sequences.'' {\it Proceedings of the IEEE}, 68.5 (1980): 593-619.
 \bibitem{pursley}M. B. Pursley, ``Performance evaluation for phase-coded spread-spectrum multiple-access communication. I-system analysis.'' {\it IEEE Transactions on Communications}, 25 (1977): 795-799.
\bibitem{KKT}W. Kuhn and A. W. Tucker, ``Nonlinear programming'', in J. Neyman (ed.), Proceedings of the Second Berkley Symposium on Mathematical Statistics and Probability (University of California Press, Berkley, CA), pp. 481-492, 1951. 
\bibitem{song_park} S. R. Park, L. Song and S. Yoon, ``A new polyphase sequence with perfect even and good odd cross-correlation functions for DS/CDMA systems.'' {\it IEEE Transactions on Vehicular Technology}, 51.5 (2002): 855-866.
\bibitem{ofdmcdma} H. Schulze, and C. L\"uders. ``Theory and applications of OFDM and CDMA: Wideband wireless communications''. John Wiley \& Sons, 2005.
\bibitem{vandercorput}J.G. Van der Corput, ``Verteilungsfunktionen I und II''. {\it Proc. K. Ned. Akad. Wet}., 38 (1935), p. 813 .
\bibitem{hansen}E. R. Hansen, ``A table of series and products.'' Prentice Hall Series in Automatic Computation, Englewood Cliffs: Prentice Hall, 1975 1 (1975), Eq. (24.1.2).
\end{thebibliography}
\end{document}